\documentclass[useAMS,usenatbib]{mn2e}

\usepackage{graphicx,amssymb,hyperref}
\usepackage[fleqn]{amsmath}  
\usepackage{multirow}
\usepackage{color}
\usepackage{hhline}
\usepackage{hyperref}
\usepackage{color,soul}

\def\mnras{Monthly Notices}
\def\simlt{\lower.5ex\hbox{$\; \buildrel < \over \sim \;$}}
\def\simgt{\lower.5ex\hbox{$\; \buildrel > \over \sim \;$}}

\def\aap{A\&A}
\def\apj{ApJ}
\def\apjl{ApJL}

\newcommand{\be}{\begin{equation}}
\newcommand{\ee}{\end{equation}}
\newcommand{\ba}{\begin{eqnarray}}
\newcommand{\ea}{\end{eqnarray}}

\title[Dark matter in galaxy cluster Abell 3827]{A test for skewed distributions of dark matter,\\ and a possible detection in galaxy cluster Abell 3827} 
\author[P.\ Taylor et al.]{Peter Taylor$^{1,2}$\thanks{e-mail: {\tt peterllewelyntaylor@gmail.com}},
Richard Massey$^{1,3}$,
Mathilde Jauzac$^{1,3,4}$, 
Fr\'ed\'eric Courbin$^{5}$, \newauthor
David Harvey$^{5}$,
R\'emy Joseph$^{5}$ \& 
Andrew Robertson$^{3}$ \\
$^{1}$ Centre for Extragalactic Astronomy, Durham University, South Road, Durham DH1 3LE, UK\\
$^{2}$ Mullard Space Science Laboratory, University College London, Holmbury St Mary, Dorking, Surrey RH5 6NT, UK\\
$^{3}$ Institute for Computational Cosmology, Durham University, South Road, Durham DH1 3LE, UK\\ 
$^{4}$ Astrophysics and Cosmology Research Unit, School of Mathematical Sciences, University of KwaZulu-Natal, Durban 4041, South Africa\\
$^{5}$ Laboratoire d'astrophysique, Ecole Polytechnique F\'ed\'erale de Lausanne, Observatoire de Sauverny, CH-1290 Versoix, Switzerland
}
\begin{document}
\date{ Accepted ---. Received ---; in original form \today}

\pagerange{\pageref{firstpage}--\pageref{lastpage}} \pubyear{2014}

\maketitle

\label{firstpage}

\begin{abstract}
Simulations of self-interacting dark matter (SIDM) predict that dark matter should lag behind galaxies during a collision.
If the interaction is mediated by a high-mass force carrier, the distribution of dark matter can also develop asymmetric dark matter tails.
To search for this asymmetry, we compute the gravitational lensing properties of a mass distribution with a free {\em skewness} parameter.
We apply this to the dark matter around the four central galaxies in cluster Abell~3827.
In the galaxy whose dark matter peak has previously been found to be offset, we tentatively measure a skewness $s=0.23^{+0.05}_{-0.22}$ in the same direction as the peak offset.
Our method may be useful in future gravitational lensing analyses of colliding galaxy clusters and merging galaxies.
\end{abstract}


\section{Introduction}

Most of the mass in the Universe is dark matter \citep[e.g.][]{planck}.
Dark matter appears invisible, because it does not interact (or interacts very weakly) with Standard Model particles including photons.

As the nature of dark matter remains unknown, there is no reason to a priori assume a particular theory of its origin.
The wide range of proposed dark matter models predict different spatial distributions, particularly on small scales.
Dark matter particles that interact with each other (SIDM) were proposed in \citep{spergel2000observational} to explain small scale discrepancies between observations and simulations of collisionless dark matter. 
In the SIDM paradigm, energy transfer between particles makes the centre of galaxies \citep{volgel12} and galaxy clusters \citep{rocha13} more circular and less dense, potentially resolving the core/cusp problem.
Small substructures can also be erased -- leading to the observed underabundance of galaxies in the Local Group, relative to simulations.
During mergers between galaxies or galaxy clusters, dark matter interactions transfer momentum between the colliding dark matter haloes \citep{randall08,kahlhoefer2014colliding,robertson2016does,kim16,robertson2017}.
These scatterings can temporarily separate dark matter from its associated galaxies. 
Such dark matter lags behind the galaxies, toward the position of diffuse gas that is slowed by ram pressure \citep{clo04,lage14,harvey2015nongravitational}. Scattering processes during collisions can be seperated into two types: frequent low momentum transfer and infrequent high momentum transfer. These will have different qualitative behaviours.

Frequent low momentum transfer scattering will cause an effective drag force, which if greater than the gravitational restoring force, will seperate the entire DM halo from the galaxy during collisions. Crucially there will be no tail of scattered DM particles escaping the potential well \citep{kahlhoefer2014colliding}. Numerous studies have placed constraints on the cross-section of DM in this regime. Measuring dark matter galaxy offsets on a sample of 72 merging clusters, \citep{harvey2015nongravitational} found  $\tilde \sigma/{m_\mathrm{DM}}< 0.5$\,cm$^2$g$^{-1}$. Constraints from the Bullet Cluster place $\tilde \sigma/{m_\mathrm{DM}}\gtrsim 1.2$\,cm$^2$g$^{-1}$  \citep{kahlhoefer2014colliding}, while  constraints from an offset galaxy in Abell 3827 yields $\tilde \sigma/{m_\mathrm{DM}}\gtrsim 2.0$\,cm$^2$g$^{-1}$ \citep{kahlhoefer2015interpretation}.

In contrast infrequent high momentum transfer scattering (mediated by a high-mass force carrier, for example) will cause a small fraction of scattered particle to leave the potential well on the trailing side. Shortly after collision, this will appear as a tail of scattered DM particles \citep[see Figure 5 in][]{kahlhoefer2014colliding}. Although the peak of the DM distribution will remain conincident with the galaxy, the tail of DM particles will lead to an apparant shift in the centre \citep{kahlhoefer2014colliding}.

Gravitational lensing offers the most direct way to map the spatial distribution of dark matter, and hence to infer its particle properties.
Gravitational lensing refers to the deflection of light rays passing near any mass, including dark matter.
Thanks to this deflection, (unrelated) objects behind dark matter appear characteristically distorted, or even visible along more than one (curved) line-of-sight.
Even though dark matter is invisible, it is possible to invert this process and infer where it must be, by undistorting the observed images, or ray-tracing multiple images back onto each other.

Galaxy cluster Abell 3827 ($22$h\,$01\arcmin$\,$49\farcs1$ $-59^\circ$\,$57\arcmin$\,$15\arcsec$, redshift $z$=0.099) is particularly well suited for this kind of study.
It gravitationally lenses a $z$=1.24 galaxy with spiral arms and several knots of star formation that can be treated as independent background sources \citep[][hereafer M15]{massey2015behaviour}.
While most clusters contain only one brightest central galaxy, Abell~3827 contains four equally-bright galaxies within its central 10\,kpc \citep{carrasco10,williams2011light}.
This highly unusual configuration means that some of the galaxies appear close to gravitationally lensed images.
Thus, under parametric model assumptions, the distribution of the dark matter can be measured.
Because of the cluster's relative proximity (in terms of gravitational lensing), it is possible to resolve small spatial offsets between the distribution of dark matter and stars in the foreground galaxies.

In this paper we present a new parametric lensing approach to search for the predicted asymmetry in the distribution of dark matter during mergers.
A previous search looked for residuals after subtracting the symmetric component \citep{harvey2017looking}, but that may be less sensitive because a tail of scattered particles shifts the best-fit position of the symmetric component backwards, thus removing some of the residual.
We instead construct a single halo model with a free {\em skewness} parameter that qualitatively captures the asymmetry found in high momentum transfer scattering simulations.
We implement and distribute this model in the publicly available {\sc Lenstool} software\footnote{\url{http://projets.lam.fr/projects/lenstool/wiki}} \citep{jullo2007bayesian}.
We test it on both mock data, where the skewness of the lens is known {\it a priori}, and on Abell~3827.
Section~2 describes existing observations of Abell~3827.
Section~3 introduces our new parametric lens model.
Section~4 contains an analysis of Abell~3827.
To be consistent with M15 we assume throughout this paper a flat $\Lambda$CDM cosmology with $\Omega_M$=0.3, $\Omega_\Lambda$=0.7 and $H_0$=70\,km\,s$^{-1}$\,Mpc\,$^{-1}$. At the redshift of Abell~3827, $1\arcsec$ corresponds to $1.828$\,kpc.

\section{Data}

\begin{figure}
\begin{center}
\includegraphics[width=\linewidth]{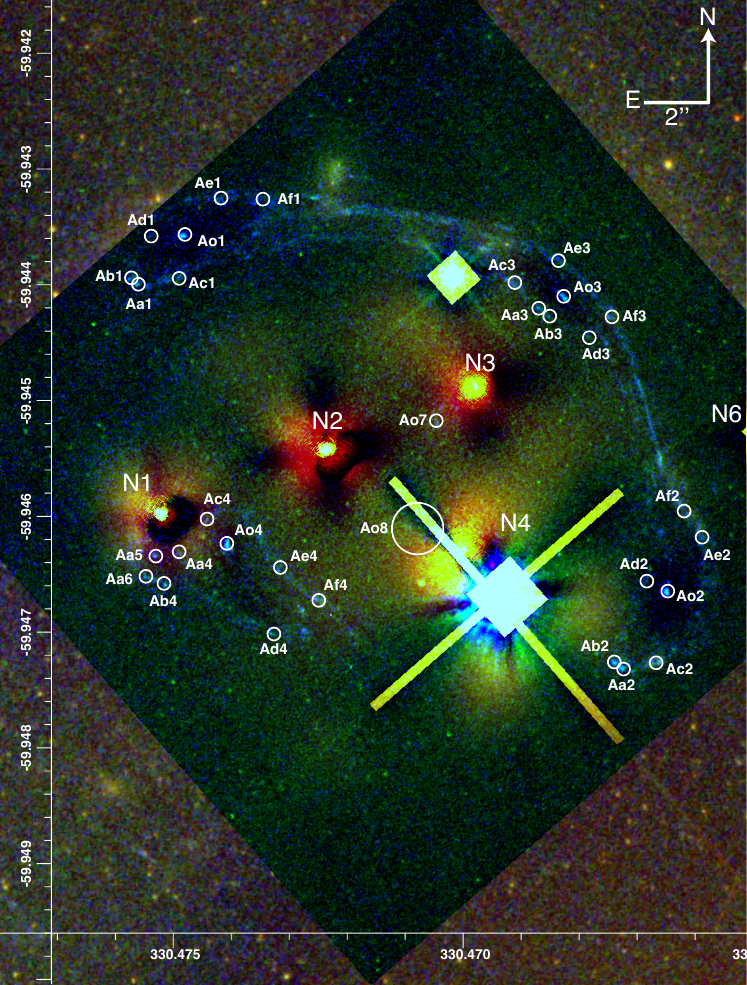}
\caption{
{\sl Hubble Space Telescope} image of galaxy cluster Abell~3827 in F160W (red), F814W (green) and F330W (blue) bands, after using {\sc MuSCADeT} \citep{joseph2016foreground} to fit and subtract foreground emission.
Residual emission from two Milky Way stars has been masked out, and remains visible at low level around the four bright central galaxies N1--N4.
Circles show multiple image identifications, with the radius of the circle reflecting uncertainty on their positions (Ao8 has only been detected from the ground).}
\label{fig:sources}
\end{center}
\end{figure}

Broad-band imaging of Abell 3827 has been obtained from the Gemini telescope at optical wavelengths \citep{carrasco10} and from the {\sl Hubble Space Telescope} ({\sl HST}) 
in the UV, optical and near-infrared (M15). 
This revealed four similarly-bright elliptical galaxies within 10~kpc of each other, plus a background spiral galaxy, 
whose multiply-lensed images are threaded throughout the cluster core (figure~\ref{fig:sources}).

Integral Field Unit spectroscopy has been obtained from the {\sc VLT}. 
An initial 1 hour exposure with the {\sl Multi-Unit Spectroscopic Explorer (MUSE)} 
identified four main groups of lensed images, and suggested two low S/N peaks as candidates for a demagnified central image (M15).
A subsequent additional 4 hour exposure (programme 295.A-5018; Massey et al.\ in prep.) confirms both candidates (Ao7 at RA: $330.47047$, Dec: $-59.945183$, Ao8 at RA: $330.47079$, Dec: $-59.946112$).
Indeed, Ao7 is also visible in {\sl HST} imaging, after using the {\sc MuSCADeT} multiwavelength method \citep{joseph2016foreground} to estimate and subtract bright foreground emission (Figure~\ref{fig:sources}).
We therefore use all the images identified by M15, plus the two new ones.

\section{Lens modelling method}

We shall model the distribution of mass in the galaxy cluster as a sum of cluster-scale plus galaxy-scale halos \citep[following e.g.][]{limousin2007truncation,jauzac14}, each a perturbation around the Pseudo Isothermal Elliptical Mass Distribution \citep[PIEMD;][]{kassiola1993elliptic}.

\subsection{Pseudo-Isothermal Skewed Potential}

The 2D surface mass density $\Sigma$ of a {\em circularly symmetric} pseudo-isothermal mass distribution, projected along a line of sight, is:
\begin{equation} \label{eq:piemd density}
\Sigma(\mathbf{r}) \equiv \frac{\sigma_0^2\,r_{\text{cut}}}{2G \left( r_{\text{cut}} - r_{\text{core}}\right)} \left( \frac{1}{\sqrt{r_{\text{core}}^2 + r^2}} - \frac{1}{\sqrt{r_{\text{cut}}^2 + r^2}} \right),
\end{equation}
where $\sigma_0$ is the 1D velocity dispersion, and where $r_{\text{core}}$ ($r_{\text{cut}}$) is an inner (outer) radius. 
To convert this into a PIEMD with ellipticity $\epsilon=\frac{a-b}{a+b}\ge 0$, where $a$ and $b$ are the semi-major and semi-minor axes respectively, \cite{kassiola1993elliptic} apply their coordinate transformation (2.3.6):
\begin{equation}
x\rightarrow x_{\text{em}}=\frac{x}{1+\epsilon} ~,~~~~~ y\rightarrow y_{\text{em}}=\frac{y} {1-\epsilon} ~.
\end{equation}
This maps a circle onto an ellipse centered at the origin, with its major axis along the $x$ axis.
In general, including a rotation to set the major axis at angle $\phi_\epsilon$, this can be expressed in polar coordinates as:
\begin{equation} \label{eqn:piemdr}
r^2 \rightarrow r_{\text{em}}^2 = \frac{r^2}{\left(1-\epsilon^2\right)^2}
\left[ 1+\epsilon^2-2\epsilon\cos{\big(2(\theta-\phi_\epsilon)\big)} \right].
\end{equation}

The angle $\alpha$ by which a light ray is deflected as it passes near the lens, and the equivalent 2D gravitational potential $\psi$ can be computed by integrating the density distribution:
\begin{equation} 
\begin{aligned}
\alpha \left( \mathbf{r}\right) = & \frac{4G}{c^2} \frac{D_l D_{ls}}{D_s} \int \hspace{-1mm} \Sigma \left( \mathbf{r'}\right) \frac{\mathbf{r}-\mathbf{r} '}{ \left| \mathbf{r}-\mathbf{r} '\right| ^ 2} ~\text{d} ^2 \mathbf{r}'\\
\label{eq:density2}
\psi \left( \mathbf{r}\right) = & \frac{4G}{c^2} \frac{D_l D_{ls}}{D_s} \int \hspace{-1mm} \Sigma \left( \mathbf{r'}\right) \text{log} \left| \mathbf{r}-\mathbf{r} '\right| ~\text{d} ^2 \mathbf{r}' ~,\\
\end{aligned}
\end{equation}
where $D_{l}$, $D_{s}$ and $D_{ls}$ are the angular diameter distance from the observer to the lens, observer to the source, and lens to the source respectively.
For general mass distributions, these integrals are difficult to solve -- but closed forms have been found for the PIEMD, using 
techniques from complex analysis that exploit its elliptical symmetry \citep{bourassa1975theory}. 

A related halo model is the Pseudo Isothermal Elliptical Potential \citep[PIEP;][]{kassiola1993elliptic}.
In this, the coordinate transformation is applied to a circular potential $\psi$ (rather than the density).
It is then mathematically easier to obtain the deflection angle and density via differentiation:
\begin{equation} \label{eq:potential}
\begin{aligned}
\alpha \left(\bold{r} \right) = & \nabla \psi \left( \bold{r}\right) \\
\Sigma \left( \bold{r} \right) = & \frac{c^2}{8 \pi G} \frac{D_s}{D_l D_{ls}} \nabla ^2 \psi \left( \bold{r} \right). \\
\end{aligned}
\end{equation}
In detail, the PIEP potential $\psi$ is transformed so that:
\begin{equation}
\psi \left( x,y \right) \rightarrow \psi ' \left( x,y \right) \equiv \psi \left( x',y' \right).
\end{equation}
The first and second derivatives can then be computed with applications of the chain rule. For example, the first $x$-derivative of the potential is:
\begin{equation}
 \psi' _x =\left( \psi_{x'}\left( x', y' \right) x'_x  + \psi_{y'} \left( x', y' \right)  y'_x \right) \big|_{\left( x,y \right)},
\end{equation}
where the subscript denotes partial differentiation.
The resulting mass distribution is not the same as a PIEMD, because of the way the coordinate transformation propagates through the chain rule (or back up the integrals in equation~\ref{eq:density2}).
For large $\epsilon$, the mass distribution corresponding to a PIEP has undesirable features including concave (peanut-shaped) isodensity contours \citep{kassiola1993elliptic}.

\subsection{Pseudo-Isothermal Skewed Mass Distribution} \label{sec:pismd}

To perturb the mass distribution in a way that resembles the behaviour of SIDM in numerical simulations \citep[see figure 5 of][]{kahlhoefer2014colliding}, we apply a further coordinate transformation that maps a circle onto an ellipse with its focus (rather than centre) at the origin:
\begin{equation}
r^2\rightarrow r^{\prime2} = \frac{r^2\left(1-s^2\right)^{3/2}}{\left(1+s\cos{[\theta-\phi_s]}\right)^2}
\end{equation}
with $s$ being the third eccentricity such that $s=\sqrt{1-b^2/a^2}$, and the power $3/2$ being introduced to preserve area. 
Note the asymmetric $\cos{(\theta)}$ terms rather than the $\cos{(2\theta)}$ terms in the mapping described by equation~\eqref{eqn:piemdr}.

\begin{figure}
\begin{center}
\includegraphics[width=\linewidth]{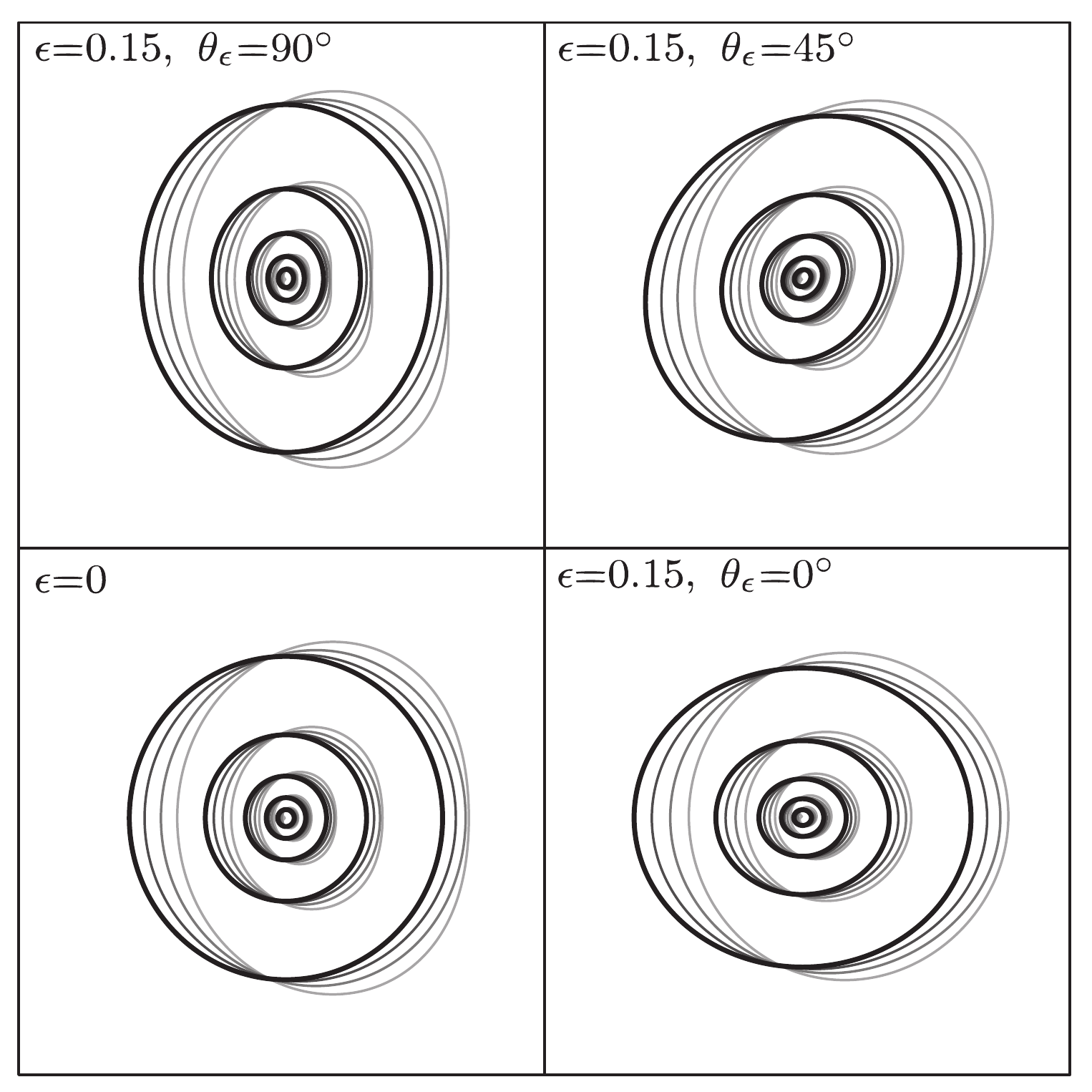}
\caption{Isodensity contours of a PISP mass distributions.
The core radius is about the same as the innermost density contour.
Thick, black lines show an ordinary PIEMD; the bottom-left is circular, and the others have ellipticity $\epsilon$=0.15, with the major axis at various angles. 
Thinner, grey lines show the same density profiles with skew $s$=0.1, 0.2, 0.3 to the right. 
}
\label{fig:coordinate_skew}
\end{center}
\end{figure}

We apply this transformation to the 2D gravitational potential corresponding to the PIEMD.\footnote{We would ideally apply this transformation to the PIEMD mass distribution, but the relevant integrals (equation~\ref{eq:density2}) do not contract to a simple form. A skewed mass distribution could also be derived from the potential corresponding to a PIEP. We choose to perturb the PIEMD so that we recover this widely-used mass distribution in the $s\rightarrow0$ limit, and to minimise undesired convex curvature in density isophotes.}
Analytic (albeit cumbersome) expressions for deflection angle and density can be readily calculated via differentiation (equation~\ref{eq:potential}).
We denote this the Pseudo Isothermal Skewed Potential (PISP); its isodensity contours are shown for various values of $\epsilon$ and $s$ in Figure~\ref{fig:coordinate_skew}. 

For any skew, the peak of the mass distribution always lies at the same position, so it will be possible to use the same metric as M15 to measure any offset between the most gravitationally bound stars and dark matter.
The total mass changes slightly with increasing skew, but this can be recalculated after a fit.

Like the PIEP, the density distribution of the PISP exhibits undesired behaviour with large skews, because the coordinate transformation was applied to $\psi$, not $\Sigma$. 
Isophotes of the density distribution become concave, and the skew ellipticity can overwhelm the underlying ellipticity.
We avoid these effects by restricting $|s|<0.3$.\footnote{See the Appendix for an alternative model that does not suffer from this effect, but has other disadvantages.}
Since the PISP is invariant under transformations $s\rightarrow -s$ and $\phi_s\rightarrow\phi_s+\pi$, we fit parameters within the domain $s\in[-0.3,0.3]$ and $\phi_s$ in some interval of length $\pi$. 
This ensures that the parameter space can be explored symmetrically about $s=0$, allowing the case of zero skew to be recovered without bias.
We set the edge of the domain of $\phi_s$ well away from any preferred direction (in practice, having explored parameter space via a quick search), to make sure an MCMC sampler operates efficiently near regions of interest.

\subsection{Testing an implementation in {\sc Lenstool}} \label{sec:tests}

We have implemented the PISP as potential 813 in the publicly-available software {\sc Lenstool} \citep{jullo2007bayesian}.
Given a parameterised mass distribution, and the location of background sources, {\sc Lenstool} can compute the position of observed multiple images. 
Given the position of observed images, it can also use Markov-Chain Monte Carlo (MCMC) optimisation to fit parameters of the lensing mass distribution.

To test whether {\sc Lenstool} can accurately recover a known input skew, we run two sets of tests. 
We first consider an isolated lens, with three background sources at different redshifts: the \textit{example\_with\_images} configuration that is packaged with {\sc Lenstool}.
As a null test, we adopt the input mass distribution with skew $s_\mathrm{true}=0$.
From the observed positions of multiple images, {\sc Lenstool} successfully recovered a best fit (maximum likelihood) value $s=-0.0008^{+0.02}_{-0.02}.$ 

We then set skew $s_\mathrm{true}=0.2$ and $\phi_{s_\mathrm{true}}=1.6$.
We set source positions by projecting one image of each source back to its source plane, then 
create a mock set of multiply lensed images by re-projecting this source forward through the lens.
When fitting this mock data, {\sc Lenstool} 
successfully recovers best-fit values $s=0.2^{+0.001}_{-0.001}$ and $\phi_s=1.6 ^{+0.04}_{-0.05}$.

Second, we test the recovery of input skews in a complex cluster lens with a mass distribution based on the Abell~3827. 
Choosing one of the quadruply lensed background galaxy images, we repeat the procedure outlined above: projecting the light backwards and then forwards through a cluster lens with known mass distribution. 
The cluster is given the same parameters as our fiducial model for Abell 3827 (see \S\ref{sec:mass_model}), with the exception of the skew parameters. 
In this test, the dark matter associated with galaxy N1 is given skewness $s_\mathrm{true}=0.25$ and $\phi_{s_\mathrm{true}}=1.6$.
As a null test, galaxies N2--N4 are set to have no skew, $s_\mathrm{true}=0$.

We run {\sc Lenstool} with the same free parameters and priors as in \S\ref{sec:mass_model}).
Within such a highly dimensional parameter space, we find that the best-fit values are sometimes noisy, for parameters that make only a small difference to the overall goodness of fit.
However, the full posterior probability distribution function (PDF) is smooth and well-sampled.
Hence, for the rest of this paper, we shall quote the modal peak and 68\% width of the posterior PDF, which {\sc Lenstool} also returns.
This makes no difference for the simple model above, and successfully recovers $s=0.24^ {+0.04}_{-0.31}$ and $\phi_s=1.6 ^{0.92}_{-0.99} $ for galaxy N1, and 
$s=0.01^{+0.14}_{-0.13}$, $0.07^{+0.10}_{-0.15}$, $0.11^{+0.11}_{-0.16}$ for galaxies N2, N3, N4 (with very weakly constrained $\phi_s$).

\subsection{Prior bias for polar parameters} \label{sec:projecting}

A skew is a two-component vector, and can be expressed in polar form as a magnitude $|s|$ and direction $\phi_s$, or in Cartesian form as an amount in  orthogonal directions $\{s_x,s_y\}$.
We implemented the polar option, so that {\sc Lenstool}'s MCMC optimiser can explore a circularly symmetric region, with no preferred direction that could bias the inferred skew.
{\sc Lenstool} also defines ellipticities in this way, for the same reason.

Nonetheless, it may often be desirable to know the posterior probability distribution of skewness along e.g.\ a direction of motion, and perpendicular to that (i.e.\ the Cartesian form).
The posterior probability distributions of skew and skew angle are returned by {\sc Lenstool} (in {\sc runmode=3}) by the sampling density of the MCMC chain.
This can be converted to the posterior of the skew in some direction $\phi$ by projecting and then weighting each sample by:
\begin{equation}
\label{eq:data weight}
w=\frac{|s|}{\sqrt{0.3^2 - s^2\mathrm{cos}^2(\phi_s-\phi)}}.
\end{equation}
The numerator is the Jacobian to convert the area of parameter space from polar to Cartesian coordinates. 
The denominator corrects for prior bias, because the restriction $|s|\in[-0.3,0.3]$ leads to a (semi-)circular prior on the projected skew.

\section{Strong lens analysis of Abell~3827}

We use the observed positions of lensed multiple images to fit a mass model of the cluster.
Our choice of model parameters and their priors is based on those of M15, with some additional degrees of freedom.
We assume $0.8\arcsec$ uncertainty on the position of lensed image Ao8, which has only been detected from the ground.
We assume $0.2\arcsec$ uncertainty on the position of all other lensed images, which are identified by {\sl HST}.

\begin{table*}
 \centering
  \caption{
Parameters of the fiducial mass model fitted by {\sc Lenstool}. 
Quantities in square brackets are fixed. 
Errors on other quantities show 68\% statistical confidence limits, marginalising over uncertainty in all other parameters.
Stellar mass components are modelled as Hernquist profiles, with a mass (computed from flux in the F606W band), scale radius and ellipticity (fitted using {\sc Galfit}; galaxy N4 is contaminated by a nearby star).
Dark matter components are modelled as PISPs, with a 1D velocity dispersion, core and cut radii, ellipticity and skewness.
Positions are given in arcseconds relative to (R.A.: $4330.47515$, Dec.: $-59.945996$), except galaxies' dark matter components, which are relative to the position of their stars.
Angles are anticlockwise from East.
  \label{tab:pots}}
  \begin{tabular}{lrr@{}lr@{}lr@{}lr@{}lr@{}lr@{}lr@{}lr@{}lr@{}lr@{}lr@{}lr@{}lr@{}lr@{}lccc}
  \hline
  \hline
 & & \multicolumn{2}{c}{$x\,$[\arcsec]} & \multicolumn{2}{c}{$y\,$[\arcsec]} & \multicolumn{2}{c}{Mass\,[$M_\odot$]} & \multicolumn{2}{c}{$r_\mathrm{sc}\,$[\arcsec]} & & & \multicolumn{2}{c}{\multirow{2}{*}{$\epsilon$~~~}} & \multicolumn{2}{c}{\multirow{2}{*}{$\phi_\epsilon\,$[$^\circ$]~}} & \multicolumn{2}{c}{\multirow{2}{*}{$s$}} & \multicolumn{2}{c}{\multirow{2}{*}{$\phi_s\,$[$^\circ$]}}\\ 
 & & \multicolumn{2}{c}{$\Delta x\,$[\arcsec]} & \multicolumn{2}{c}{$\Delta y\,$[\arcsec]} & \multicolumn{2}{c}{$\sigma_\mathrm{v}\,$[km/s]} & \multicolumn{2}{c}{$r_\mathrm{core}\,[\arcsec]$} & \multicolumn{2}{c}{\!\!$r_\mathrm{cut}\,[\arcsec]$\!\!}
 \\
\hline
N1 & {\rm stars}  & [$-0.06$ & ] & [$0.04$ & ] & \multicolumn{2}{c}{$[1.00 \times 10 ^ {11}]$} & \multicolumn{2}{c}{[0.53]} & & & [$0.12$ & ] & [$61$ & ]
\\
\multicolumn{2}{r}{{\rm ~dark matter}} & $-0.29$ & $^{+0.25}_{-0.14}$ & $-0.71$ & $^{+0.30}_{-0.16}$ & ~~$149$ & $^{+8}_{-12}$ & \multicolumn{2}{c}{[$0.1$]} & \multicolumn{2}{c}{[$40$]} & $0.02$ & $^{+0.33}_{-0.01}$ & $151$ & $^{+19}_{-116}$ & $0.21$ & $^{+0.06}_{-0.22}$ & $86$ & $^{+44}_{-44}$ \vspace{2mm}
\\
N2 & {\rm stars} & [$5.07$ & ] & [$2.05$ & ] &\multicolumn{2}{c}{$[2.46 \times 10 ^ {11}]$} & \multicolumn{2}{c}{[0.79]} & & & [$0.17$ & ] & [$39$ & ]
\\
\multicolumn{2}{r}{{\rm dark matter}} & $-0.23$ & $^{+0.30}_{-0.16}$ & $0.00$ & $^{+0.30}_{-0.30}$ & $182$ & $^{+29}_{-22}$ & \multicolumn{2}{c}{[$0.1$]} & \multicolumn{2}{c}{[$40$]} & $0.42$ & $^{+0.05}_{-0.22}$ & $23$ & $^{+32}_{-12}$ & $0.03$ & $^{+0.11}_{-0.14}$ & $117$ & $^{+41}_{-80}$ \vspace{2mm}
\\
N3 & {\rm stars} & [$9.69$ & ] & [$3.98$ & ] &\multicolumn{2}{c}{$[2.77 \times 10 ^ {11}]$} & \multicolumn{2}{c}{[0.33]} & & & [$0.05$ & ] & [$31$ & ]
\\
\multicolumn{2}{r}{{\rm dark matter}} & $-0.05$ & $^{+0.25}_{-0.25}$ & $-0.06$ & $^{+0.18}_{-0.29}$ & $213$ & $^{+8}_{-10}$ & \multicolumn{2}{c}{[$0.1$]} & \multicolumn{2}{c}{[$40$]} & $0.49$ & $^{+0.01}_{-0.16}$ & $15$ & $^{+14}_{-8}$ & $-0.02$ & $^{+0.08}_{-0.11}$ & $169$ & $^{+7}_{-109}$ \vspace{2mm}
\\ 
N4 & {\rm stars} & [$9.26$ & ] & [$-1.08$ & ] &\multicolumn{2}{c}{$[2.08 \times 10 ^ {11}]$} & \multicolumn{2}{c}{[1.37]} & & & [$0.39$ & ] & [$127$ & ]
\\
\multicolumn{2}{r}{{\rm dark matter}} & $-1.35$ & $^{+0.39}_{-0.34}$ & $0.51$ & $^{+0.35}_{-0.27}$ & $255$ & $^{+8}_{-10}$ & \multicolumn{2}{c}{[$0.1$]} & \multicolumn{2}{c}{[$40$]} & $0.02$ & $^{+0.25}_{-0.01}$ & $136$ & $^{+17}_{-28}$ & $0.08$ & $^{+0.08}_{-0.09}$ & $147$ & $^{+21}_{-80}$ \vspace{2mm}
\\
N6 & {\rm stars} & [$18.54$ & ] & [$2.47$ & ] &\multicolumn{2}{c}{[0]}
\\
\multicolumn{2}{r}{{\rm dark matter}} & [$0$ & ] & \multicolumn{2}{c}{[0]} & $38$ & $^{+26}_{-25}$ & \multicolumn{2}{c}{[$0.1$]} & \multicolumn{2}{c}{[$40$]} & [$0$ & ] & [$0$ & ] & \multicolumn{2}{c}{[$0$]} & \multicolumn{2}{c}{[$0$]} \vspace{2mm}
\\
\multicolumn{2}{l}{Cluster ~ {\rm dm}} & $5.53$ & $^{+1.46}_{-1.61}$ & $2.33$ & $^{+1.97}_{-1.59}$ & $683$ & $^{+139}_{-75}$ & \!\!$30.12$ &  $^{+9.23}_{-6.43}$\!\! & \multicolumn{2}{c}{[$1000$]} & $0.56$ & $^{+0.13}_{-0.10}$ & $63$ & $^{+2}_{-3}$ & \multicolumn{2}{c}{[$0$]} & \multicolumn{2}{c}{[$0$]}
\\
\hline\hline
  \end{tabular}
\end{table*}

\subsection{Fiducial mass model}
\label{sec:mass_model}

The cluster's large-scale mass distribution is modelled as a single PIEMD.
Based on a comprehensive (but slow) initial exploration of parameter space, its position is given by a broad Gaussian prior with $\sigma=2\arcsec=3.66\,\text{kpc}$, centred on the position of galaxy N2.
Flat priors are imposed on its ellipticity ($\epsilon<0.75$), core size ($r_\mathrm{core}<40\arcsec$) and velocity dispersion ($300<$$\sigma_\mathrm{v}$$<1000$\,km/s).
Its cut radius is fixed at $r_\mathrm{cut}=1000 \arcsec$, well outside the strong lensing region, i.e. away from any multiple image constraints.

Central galaxies N1--N4 are each modelled as a stellar component (which was not included in the fiducial model of M15), plus a dark matter one.
Following \cite{giocoli2012moka}, the stellar components are modelled with \cite{hernquist1990analytical} profiles:
\begin{equation}
\rho_{\text{star}} (r) = \frac{\rho_s}{\left( r/r_s \right) \left( 1+r/r_s\right)^3},
\end{equation}
where the scale radius $r_s$ is related to the half mass radius $R_e$, such that $R_e=r_s/0.551$, and the scale density $\rho_s=M_{\mathrm{total}}/\left(2 \pi r_s^3\right)$.
We fix the mass of the stellar component, and its half-mass radius, using the optical magnitudes and profiles measured by M15. 
These parameters are listed in Table~\ref{tab:pots}.

The four central galaxies' dark matter components are now modelled as PISPs.
We impose flat priors on their positions, in $4 \arcsec \times 4 \arcsec$ boxes centred on their luminosity peaks, plus flat priors on their ellipticity ($\epsilon<0.5$) and velocity dispersion ($v_\mathrm{disp}<600$\,km/s).
We fix $r_\mathrm{cut}=40\arcsec=73$\,kpc \citep{limousin2007truncation}.

Galaxy N6 is much fainter than the others, so we approximate its total mass distribution as a single PIEMD.
This has a fixed position and ellipticity to match the light distribution, and only its velocity dispersion is optimised (with a flat prior $v_\mathrm{disp}<500$\,km/s).

\begin{figure}
\centering
\vspace{2mm}
\includegraphics[width=85mm]{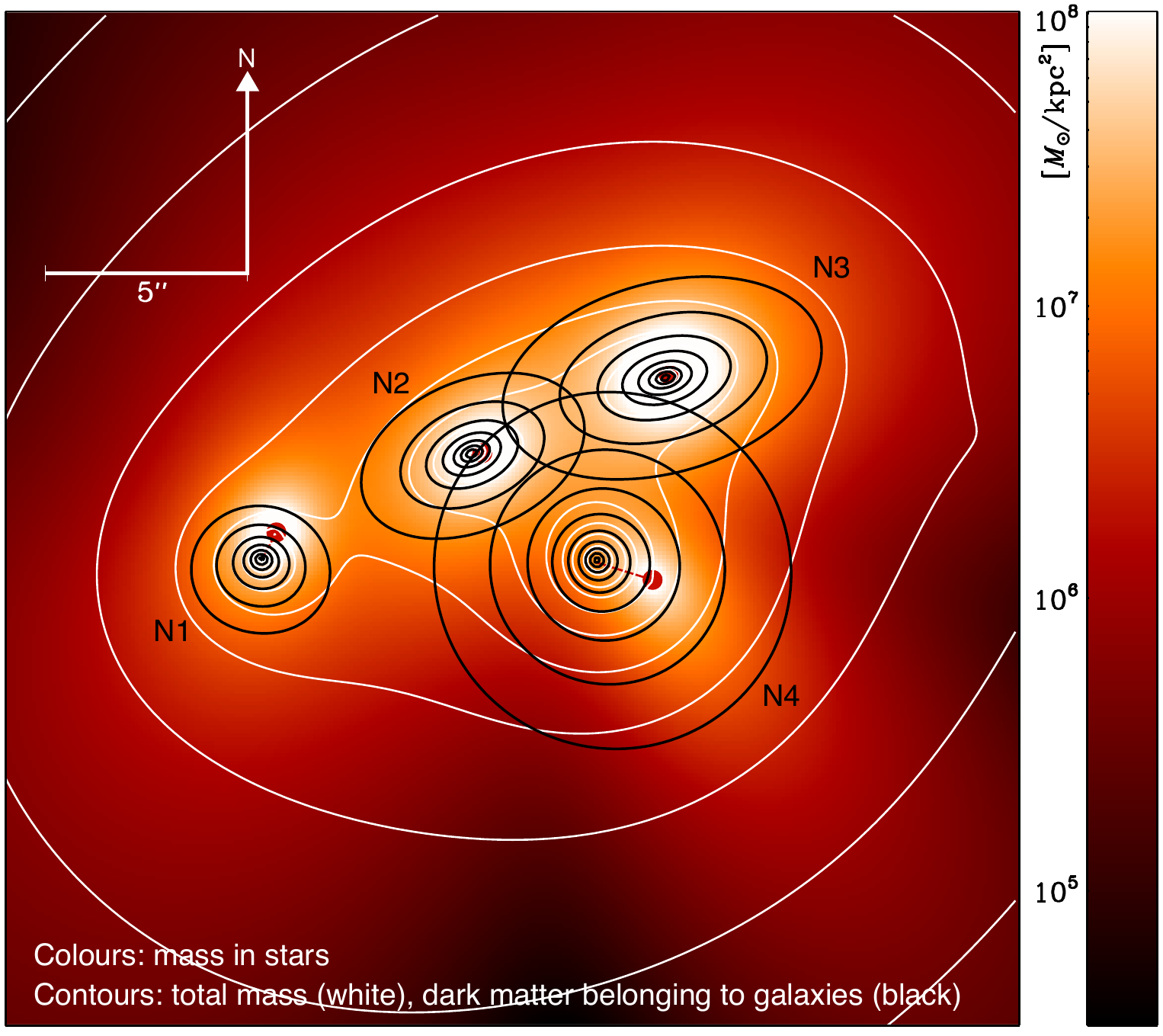}
\caption{The best fitting mass distribution in the gravitational lens Abell~3827, integrated along our line of sight.
For reference, the background colour scale shows the modelled stellar mass density. 
Red spots indicate the position of the luminosity peak in galaxies N1--N4.
White isodensity contours show the total lensing mass 
of the cluster.
The outermost contour corresponds to a projected density of $2\times10^9\,M_\odot/\text{kpc}^2$, and values increase towards the centre by a factor of $2^{1/3}$=1.26.
Black isodensity contours isolate each galaxy's dark matter component.
The outermost contour corresponds to a projected density of $1.26\times10^{9}\,M_\odot/\text{kpc}^2$ and values increase by a factor of $2^{2/3}$.
The visible offset between stars and dark matter in galaxies N1 and N4 are both statistically significant; the asymmetry in the distribution of N1's dark matter is also significant.}
\label{fig:massmap}
\end{figure}

\begin{figure*}
\includegraphics[width=83mm]{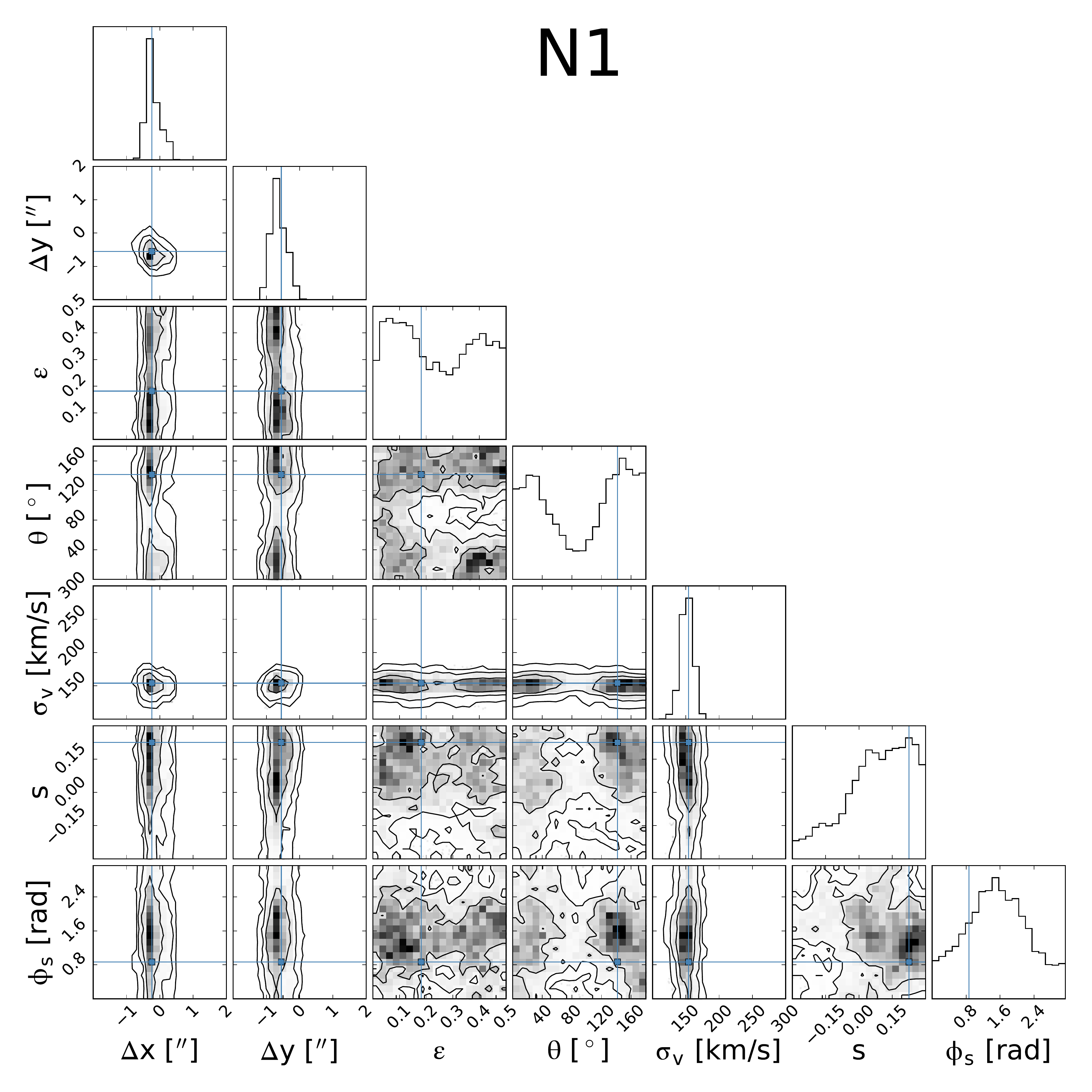}~~~
\includegraphics[width=83mm]{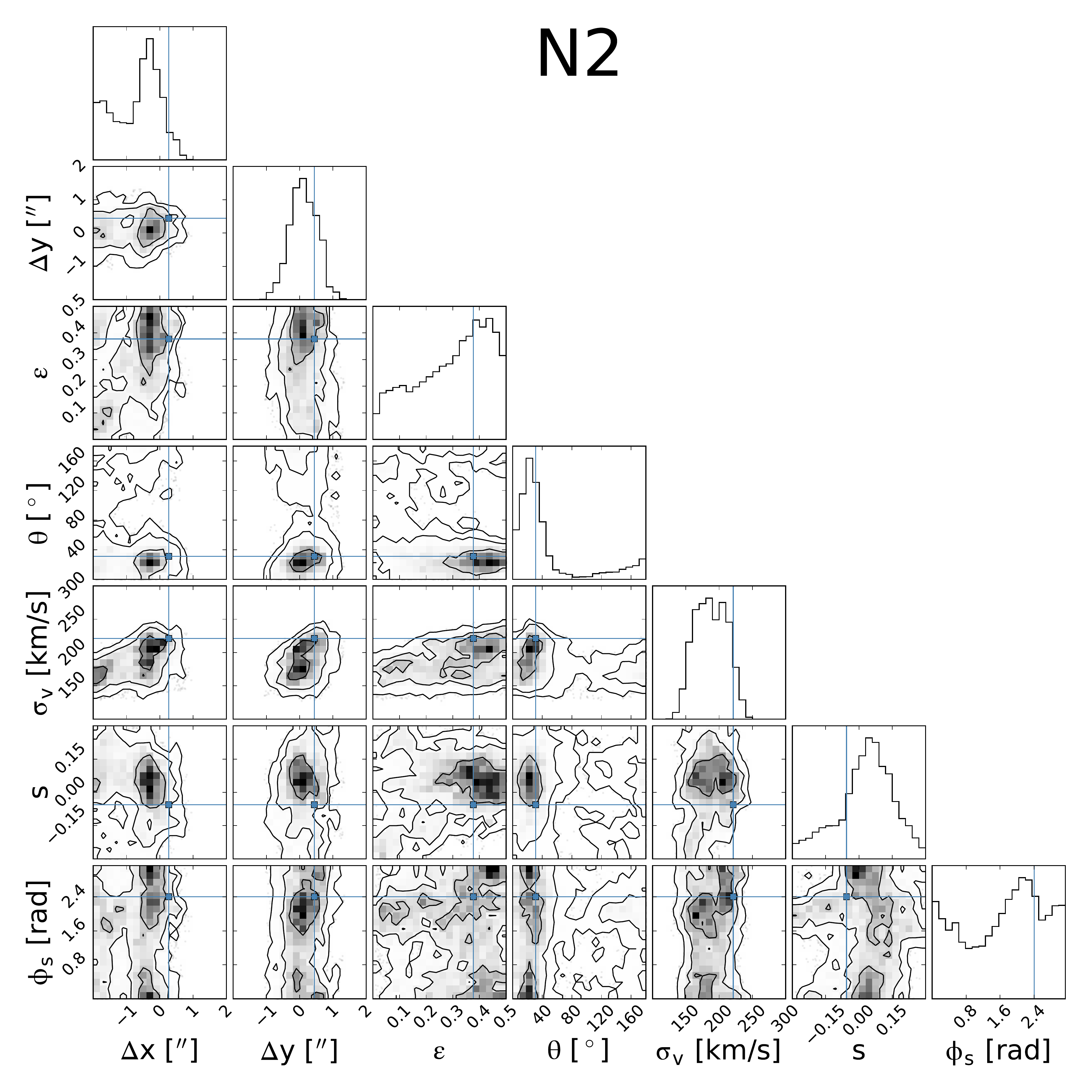}\\
\includegraphics[width=83mm]{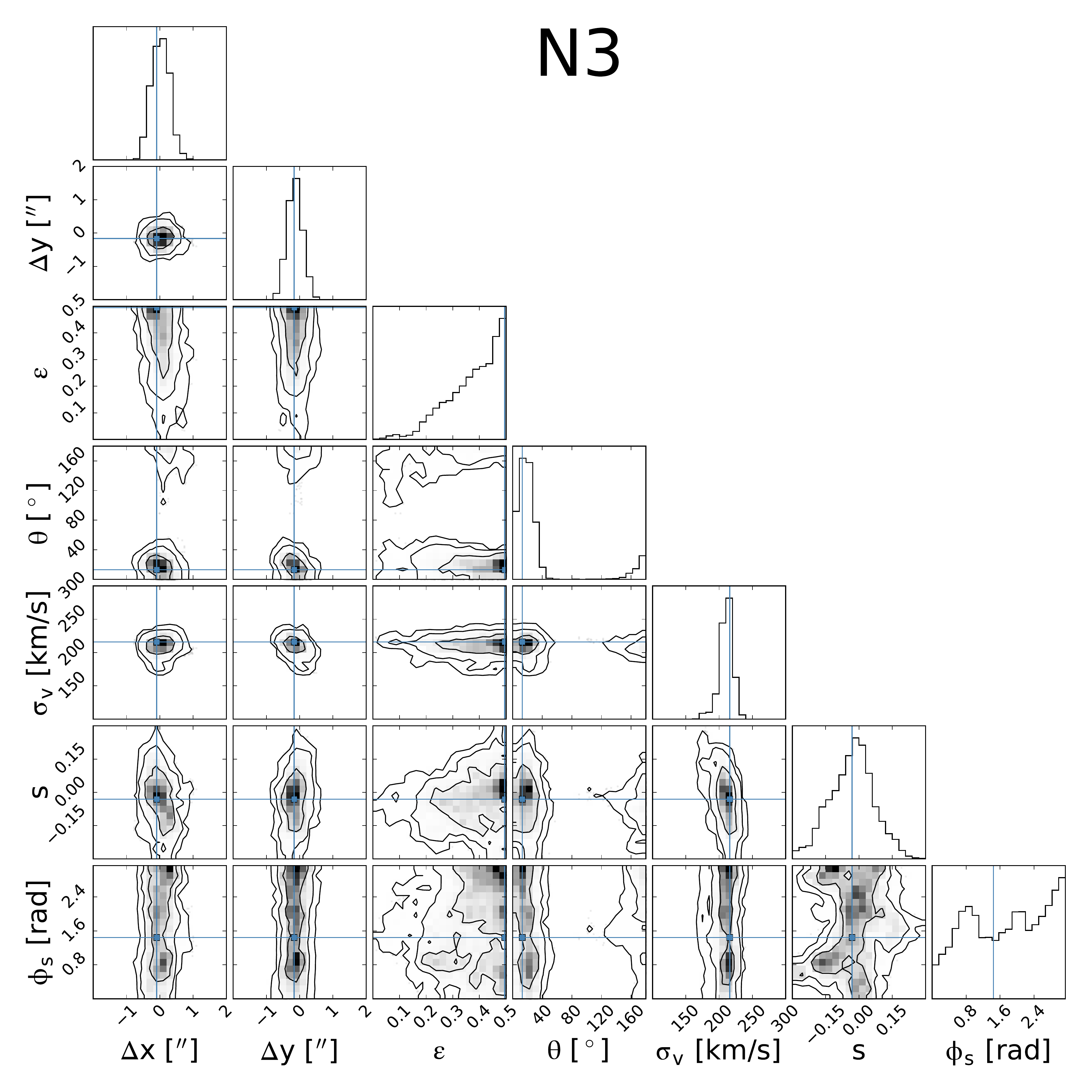}~~~
\includegraphics[width=83mm]{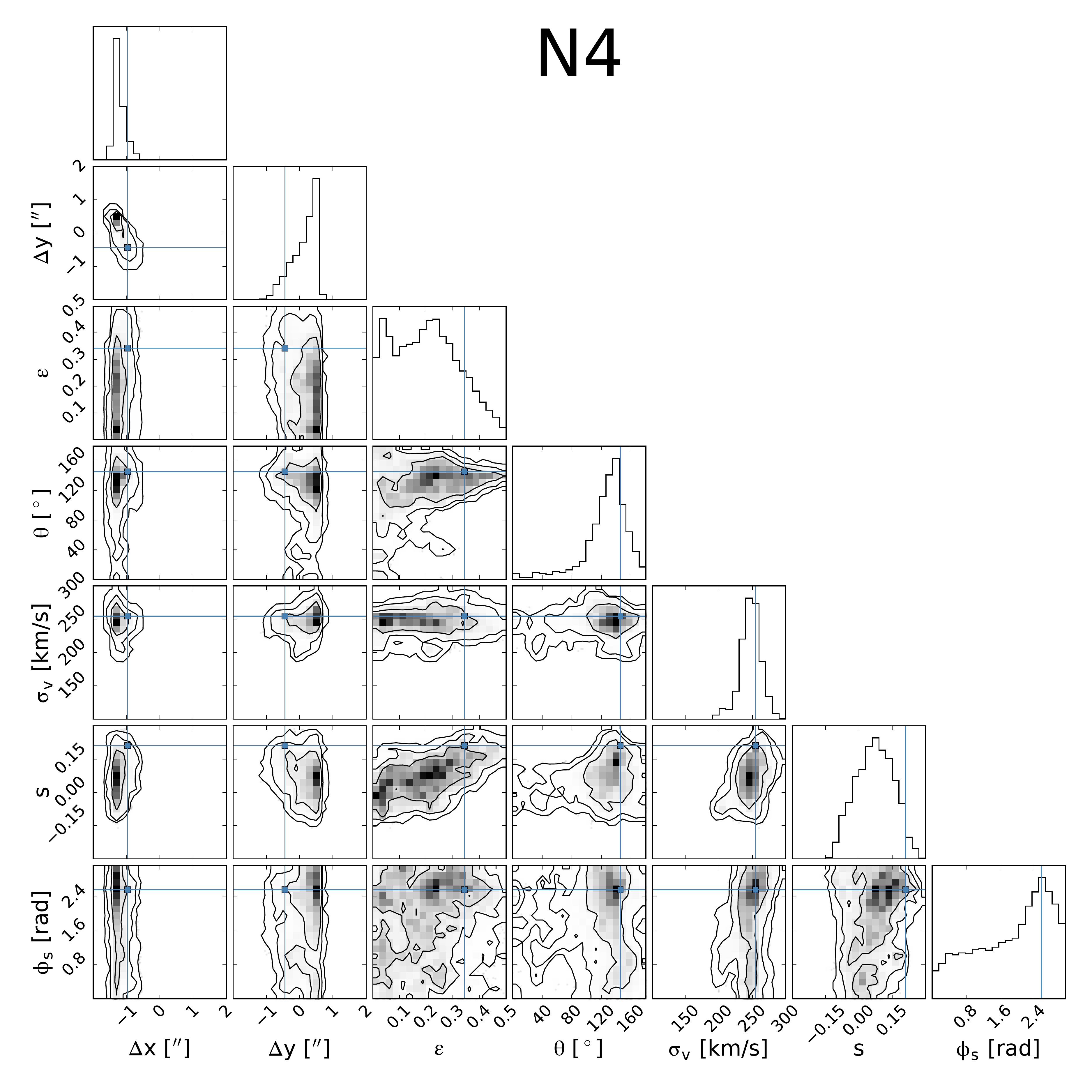}
\caption{Posterior probability distribution of the distribution of dark matter associated with galaxies N1 (top left), N2 (top right), N3 (bottom left) and N4 (bottom right).
Contours show the $68\%$, $95 \%$ and $99.7 \%$ contour levels. 
Blue lines indicate the best fit model, but note that this can be noisy, and we instead use the  peak of the smoothed posterior distribution.
Positions have been recentered such that $(x,y)=(0,0)$ is the peak of the stellar luminosity.
Offsets between stars and dark matter are measured at $>$$3\sigma$ for galaxies N1 and N4.
A skew is detected at $>$$1\sigma$ for galaxy N1, in a direction consistent with the spatial offset.}
\label{fig:gals_corner}
\end{figure*}

\begin{figure*}
\vspace{-3mm}
\includegraphics[width=160mm]{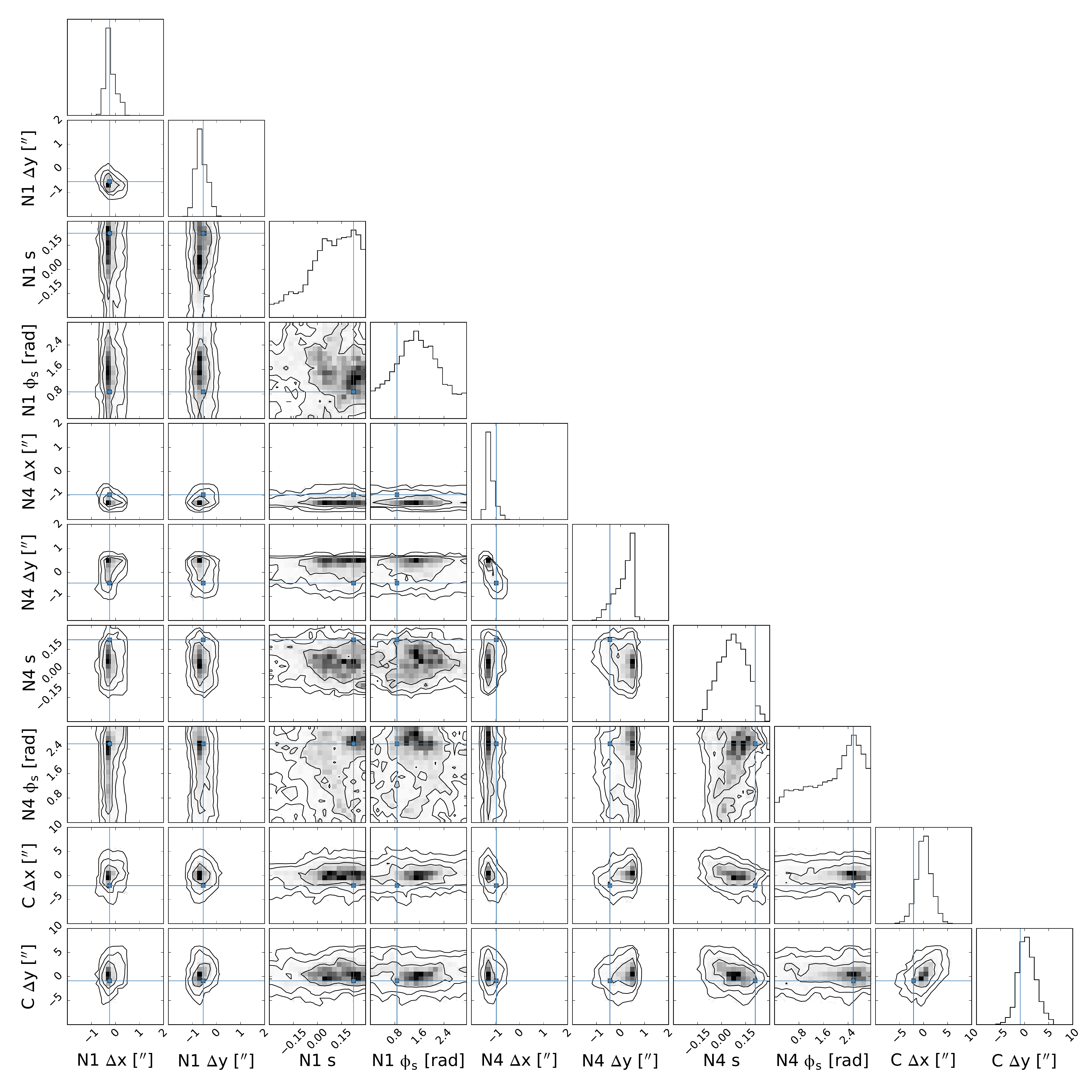}
\caption{Posterior probability distribution showing (minimal) correlations between the position and asymmetry of the dark matter associated with key galaxies, and between those galaxies and the cluster-scale dark matter(denoted C). Contours show the $68 \%$, $95 \%$ and $99.7 \%$ contour levels.
Positions have been recentered as in Figure~\ref{fig:gals_corner}.}
\label{fig:positions_corner}
\end{figure*}

We optimise the free parameters using {\sc Lenstool}, with {\sc runmode}=3. This runmode is used to fully explore the posterior \citep{jullo2007bayesian}.
(Modal) maximum likelihood parameters are shown in Table~\ref{tab:pots}, and the corresponding mass distribution is shown in Figure~\ref{fig:massmap}.
The best fit model achieves a $RMS$ offset between the observed and predicted positions of multiple images of $\left<\mathrm{rms}\right>_i$=$0.26\arcsec$. There are $54$ constraints and $35$ free parameters in our model. The modal $\chi^2/\text{dof}$=$67.1/19$ with a log likelihood of 8.18.  
The full posterior probability distribution for the dark matter associated with galaxies N1--N4 is shown in Figures~\ref{fig:gals_corner} and \ref{fig:positions_corner}.

\subsection{Sensitivity to model choices}

\subsubsection{Stellar mass components}

Galaxies definitely contain stars, and those stars have mass. 
Not accounting for this mass could bias the skew measurement. In an offset DM halo, the stellar mass will lead the DM peak, while any tail of DM particles will trail behind in the opposite direction. Not accounting for this stellar mass could weaken the skew measurement, or in the worst case scenario, if the stellar mass is greater than that of scattered particles, reverse the direction of the measured skew. We have explicitly modelled the stellar mass seperately to avoid any bias in the inferred skew.
In practice, as in M15, we find that including the stellar mass component (or even multiplying/dividing its mass by a factor 2) does not change any other results, within their statistical errors.

\subsubsection{Identification of new lensed images}

Adding constraints from the two new lensed images Ao7 and Ao8 tightens constraints on nearby galaxies N3 and N4.
These (demagnified) images are unresolved, and any of the features in the background spiral could be assigned to them.
We have tried relabelling one or both of the demagnified images as either the bulge, Ao$n$, or one of the two brightest knots of star formation, Aa$n$ or Ab$n$. 
{\sc Lenstool}'s outputs are statistically indistinguishable.
In all cases, the {\em entire} background spiral galaxy is predicted to be lensed onto both the northern and the southern demagnified images.

\subsubsection{Mass in other cluster member galaxies}

We also tested the impact of adding more cluster member galaxies to the mass model.
These galaxies were identified using a colour-magnitude selection using the F814W and F606W2 HST/ACS band imaging. 
Source detection was done using \textsc{sextractor} \citep{bertin1996} in dual mode, with reference taken in the F814W-band.
We then identified as cluster members all galaxies brighter than $mag_\mathrm{F814W} < 23$ and within 1$\sigma$ of the red sequence best-fit: 
$$
(mag_\mathrm{F814W}-mag_\mathrm{F606W}) = 0.022 \times mag_\mathrm{F606W} - 1.129 . 
$$
Our final cluster member catalogue contains 147 galaxies.

These galaxies are added to the mass model as small scale perturbers. We assume fixed cut radius and velocity dispersion, scaled by their luminosities in the F814W-band. 
This methodology has been successfully validated by \cite{harvey2016systematic}, and adopted widely in previous work \citep[e.g.][]{jauzac14,richard14,limousin07}. To derive $L^{\ast}$, we use the $K^{\ast}$ magnitudes obtained by \cite{lin2006} as a funciton of cluster redshift. \textsc{lenstool} is then scaling the cut radius and velocity dispersion of each galaxies in our catalogue relative to a $K^{\ast}=16.6$ galaxy with velocity dispersion $\sigma^{\ast} = 108.4 \pm 27.5$~km/s, and cut radius $r_\mathrm{cut}^{\ast}=48.5 \pm 16.0$~kpc.

Including all cluster memeber galaxies in a re-optimised mass model significantly affects neither offset nor skewness measurements of dark matter associated with central galaxies N1--N4. 
By far the most affected measurement is the skewness of galaxy N1, which increases to $s=0.28^{+0.01}_{-0.31}$.
All other quantities remain consistent within random noise.

\section{Discussion}

\subsection{Galaxy N1}

The previously-detected offset between galaxy N1's stars and dark matter persists at $>3\sigma$ in our new analysis.
Adding two free parameters for the asymmetry of its dark matter slightly increases uncertainty in its position.
The modal offset is $\left( \Delta x, \Delta y \right) = \left(-0.22 ^ {+0.25}_{-0.14} , -0.81 ^ {+0.16}_{-0.17} \right)$ for an unskewed model, and $\left( \Delta x, \Delta y \right) = \left(-0.29 ^ {+0.25}_{-0.14} , -0.71 ^ {+0.30}_{-0.16} \right)$ if skewness is allowed. The consistency between these suggests that the measured position of the density peak is robust against the skew parameter probed here.

If the offset is entirely due to an effective drag force through frequent dark matter self-interactions, it implies a momentum-transfer interaction cross-section $\tilde \sigma/{m_\mathrm{DM}}\gtrsim1$\,cm$^2$g$^{-1}$, assuming galaxy N1 is falling into the cluster for the first time \citep{kahlhoefer2015interpretation}.
In general, We agree with this interpretation\footnote{We have also repeated \cite{kahlhoefer2015interpretation}'s calculation of $\tilde \sigma/{m_\mathrm{DM}}$ but integrating the effect of the restoring force on the {\em entire} distribution rather than just the peak. The difference is not significant.}, but note that the cross-section can be lower if the galaxy has completed multiple orbits; its current direction of motion is unknown.

\begin{figure}
  \centering
  \includegraphics[width=80mm]{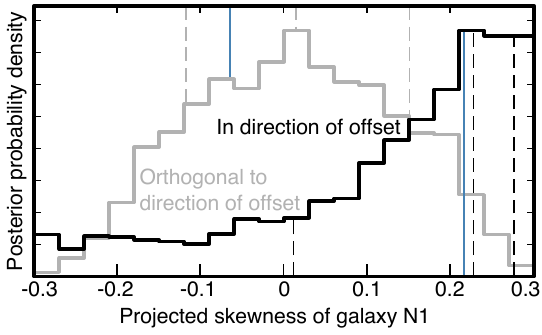}
\caption{The posterior of the skew vectors in galaxy N1 projected onto the offset vector (black) and orthogonal to this (grey). The Jacobian has been accounted for and the priors have been adjusted as described in section~\ref{sec:projecting}. Dashed lines indicate the posterior peak and $1\sigma$ confidence intervals. The blue line indicates the (noisy) best fit value. There is a preference for a skew that is consistent with SIDM. No such preference is shown for a skew component in the orthogonal direction.}
\label{fig:gal_1_skews}
\end{figure}

We also find (at much lower statistical significance) that galaxy N1's dark matter is skewed in a direction consistent with the SIDM interpretation of its offset.
This could be signs of a tail of scattered DM particles and would favour high momentum transfer scattering. However, the weak statistical significance of our result makes it impossible to rule out the low momentum transfer case.
Figure~\ref{fig:gal_1_skews} shows the posterior PDF of the skew onto the vector pointing from the DM peak to the stellar luminosity in the fiducial model, such that a positive skew corresponds to the direction predicted by SIDM. 
The peak of the posterior and $1\sigma$ errors are $s=0.23^{+0.05}_{-0.22}$.
If we individually project the skewness onto the offset direction individually in all MCMC samples, we find $s=0.26^{+0.03}_{-0.22}$. 
Finally in our model that contained the additional 147 cluster galaxies we find $s = 0.28 ^ {+0.01} _{-0.31}.$
In all cases $\sim70\%$ of the posterior PDF lies at $s>0$.
As the posterior peak is near the edge of the prior, which is chosen coservatively (see \S\ref{sec:pismd}), a parametric halo model that does not break down for large skew parameters could result in a stronger detection.

\subsection{Galaxies N2 and N3}

The dark matter associated with galaxies N2 and N3 appears symmetric, and coincident with the stars.
This result does not preclude offsets from existing along the line of sight.
Furthermore, even if the galaxies are moving in the plane of the sky, they could be behind or in front of the most dense regions of the cluster core, and therefore passing through a less dense medium, experiencing less drag.

\subsection{Galaxy N4}

The measured position of N4's dark matter is intriguing.
Accounting purely for statistical error bars, thanks to the confirmed positions of demagnified images, we find that galaxy N4 is offset from the galaxy's stars at the $3 \sigma$ level.
However, the offset position is mildly degenerate with the position of the cluster-scale dark matter (Figure \ref{fig:positions_corner}), thus a flat prior on the cluster scale halo could lead to a different offset measurement.
Furthermore, the measured ellipticity of the galaxy light is contaminated by light from an adjacent Milky Way star, and its position may also be.

Despite the measured offset of dark matter from galaxy N4, it shows no sign of skewness. If the offset is spurious, as discussed above, then 
a tail is not expected. If the offset is real, the lack of skew favours low momentum transfer scattering and (at very low S/N) is in mild tension with the skew detected in galaxy N1. Unknown systematics in the modelling of DM around either galaxy could be responsible. Nonetheless, there is also a possible physical explanation for this discrepancy. Galaxy N4 is in a higher density environment than galaxy N1, closer to the cluster core. It is possible that any tail of scatttered N4 particles has been tidally destroyed by the steeper gradient in gravitational potential.

\section{Conclusion}

We have developed  a parametric lens models for asymmetrically skewed mass distributions.
This can be used to search for scattered (self-interacting) dark matter in colliding systems.
More generally, it will also be useful to investigate claims of dynamically-induced asymmetry \citep{prasad16,chemin16}, or tidal tails (which are asymmetric if the size of a body is large compared to its distance from the centre of potential).

We have also presented a new model for the distribution of mass in galaxy cluster Abell 3827. 
Our choice of flat priors for the position of \textit{all} galaxies' dark matter leads to a detected offset between a second galaxy's stars and its dark matter. 
New VLT/MUSE observations tighten the constraints on that offset.
Neither measured offset changes significantly if the models are allowed extra freedom to become skewed.

We find tantalising, but low significance evidence that the galaxy closest to multiply lensed images (and therefore the best constrained) has an asymmetric distribution of dark matter, skewed in the same direction as its offset from stars.
We emphasise that our skew model, which captures the qualitative behaviour of scattered DM particles, is primarily motivated by mathematical convenience and that all skew measurements here are model dependent.
More work will be needed to determine the significance of this result: whether it is physical, or an artefact of systematics in parametric lens modelling. 
Even in mock data where the true skew is known, skewness cannot be measured to high precision in a system as complex as Abell~3827. 
This is probably because the effect of skewness on lensed image positions is smaller than the effects of other free parameters.

A promising direction for future investigation may be provided by pairs of field galaxy in the SLACS survey, one of which has already been found to have an offset between dark and luminous matter \citep{shu2016kiloparsec}. 
Whilst the SIDM model predicts a largest (most easily observable) offset in galaxies moving through a dense intracluster medium, it may be possible to more tightly constrain any asymmetry of dark matter in these simpler systems. 
If the directions of their dark matter tails correlate with the directions of their offsets, this evidence would support the hypothesis of SIDM with a high-mass mediator particle.

\section*{Acknowledgements}

The authors are grateful for constructive conversations with Benjamin Cl\'ement, Eric Jullo, Felix Kahlhoefer, Thomas Kitching, Marceau Limousin and Subir Sarkar.
We would also like to thank the anonymous referee whose comments have significantly improved the paper.
Figure~\ref{fig:gals_corner} was produced using the python module {\sc corner} \citep{corner}.
PT is supported by the STFC.
RM is supported by the Royal Society.
AR and MJ are supported by the UK Science and Technology Facilities Council (grant numbers ST/H005234/1 and ST/L00075X/1).

\noindent {\it Facilities:}
This paper uses data from observations GO-12817 (PI:~R.\,Massey) with the NASA/ESA {\sl Hubble Space Telescope}, obtained at the Space Telescope Science Institute, which is operated by AURA Inc, under NASA contract NAS 5-26555.
This paper also uses data from observations made with ESO Telescopes at the La Silla Paranal Observatory under programmes 294.A-5014 and 295.A-5018 (PI:~R.\,Massey).
We thank the Paranal Science Operations team for carrying out those observations.
Our {\sc Lenstool} optimisation used the DiRAC Data Centric system at Durham University, operated by the Institute for Computational Cosmology on behalf of the STFC DiRAC HPC Facility (\url{www.dirac.ac.uk}).
This equipment was funded by BIS National E-infrastructure capital grant ST/K00042X/1, STFC capital grant ST/H008519/1, STFC DiRAC Operations grant ST/K003267/1 and Durham University. 
DiRAC is part of the UK National e-Infrastructure. 

\bibliographystyle{mnras}
\bibliography{bibtex.bib}

\appendix

\section{Alternative method to generalise lens mass distributions}

Another way to introduce asymmetry is to apply a weighting function $w \left( \bold{r};\{ a_i \} \right)$ to an elliptical lensing potential
\begin{equation}
\psi(\bold{r}) \rightarrow \psi ' (\bold{r}) \equiv w (\bold{r};\{ a_{i}\}) \, \psi(\bold{r}),
\end{equation}
where $\{ a_i\}$ are a set of parameters. The deflection and surface mass density are readily computed by differentiating.

We consider weighting functions of the form
\begin{equation}
w (\bold{r}; \{a_i\}) = 1 + s f(r,\theta)
\end{equation}
where $s$ is a skew parameter and $f(r, \theta)$ is written in polar coordinates. The second term acts as a perturbation away from elliptical symmetry of $O(s)$, with $s=0$ corresponding to the elliptically symmetric case. We chose $f(r,\theta)$ to meet the following criteria:
\begin{itemize}
\item To ensure that the space about $s=0$ is explored symmetrically in Lenstool, so that a non-zero skew is not artificially recovered, we require that $sf(r,\theta) = -s f(r,\theta+\pi)$.
\item To avoid difficulties near the origin, we require  $f(r, \theta)$ to be an increasing function of $r$. This is also physically motivated, as it is difficult to scatter particles from the centre of the potential well at $r=0$.
\item To ensure that the surface mass density remains positive (or becomes negative only for large $r$ well outside any region of interest), we require $f \left(r, \theta \right)$ to be bounded.
\end{itemize}

\begin{figure}
\begin{center}
\includegraphics[width=\linewidth]{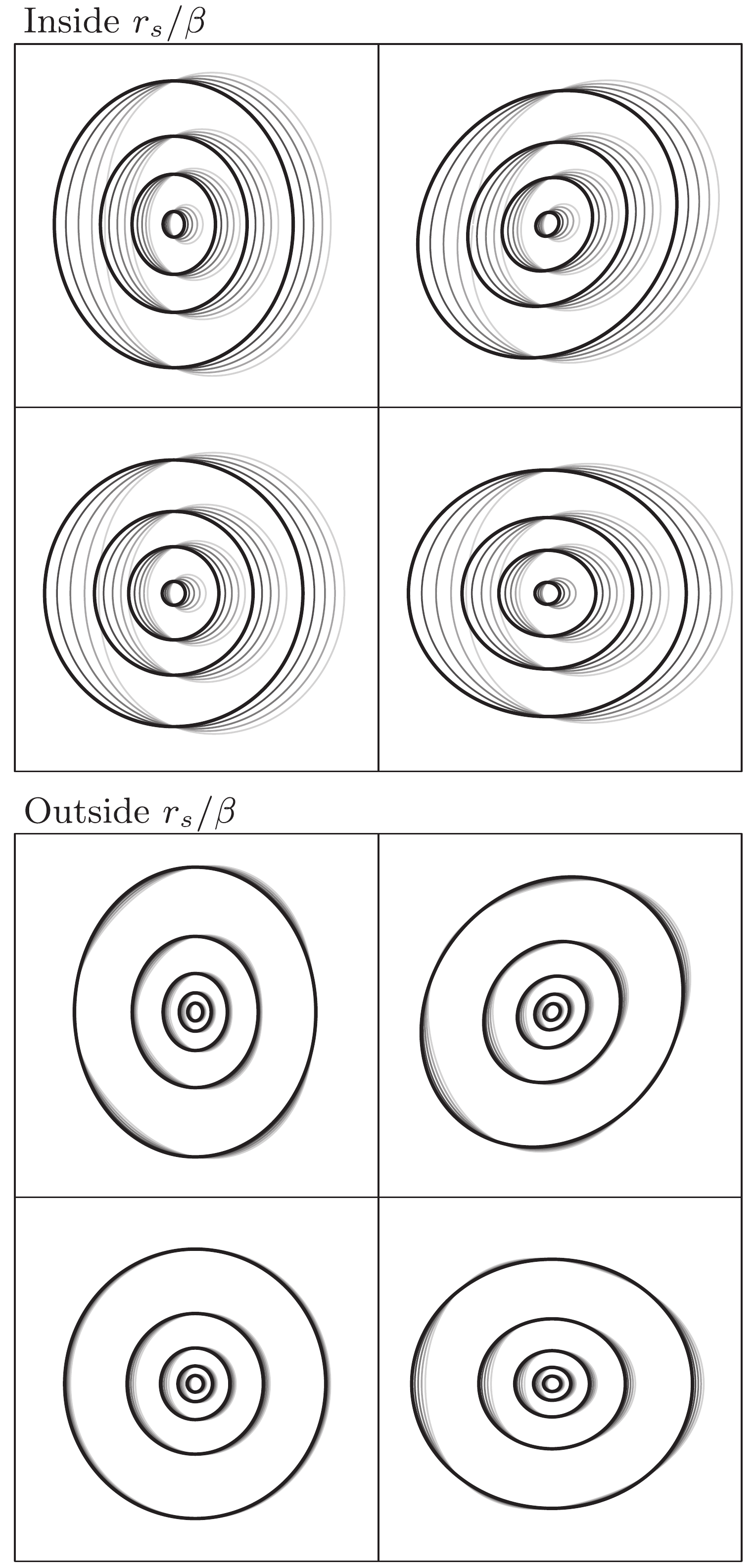}
\caption{Isodensity contours of wPISP mass distributions, inside (top) and outside (bottom) the scale radius $r_s/\beta$.
Thick, black lines show an ordinary PIEMD; the bottom-left is circular, and the others have ellipticity $\epsilon$=0.15, at angles $\theta_\epsilon$=$0^\circ$, $45^\circ$ and $90^\circ$ in the same order as in Figure~\ref{fig:coordinate_skew}.
Thinner, grey lines show the same density profiles with skew $s$=0.1, 0.2, 0.3, 0.4 to the right.
}
\label{fig:weight_skew}
\end{center}
\end{figure}

\subsection{The Weighting Function Pseudo-Isothermal Skewed Potential}

Meeting the above conditions we form the weighted Pseudo-Isothermal Skewed Potential (wPISP) by applying the weighting function
\begin{equation} \label{eq:weighting function}
w (\bold{r}; s, r_s, \beta, \phi_s) = 1 + s \text{ tan}^{-1} \left(\beta \frac{r}{r _{s}}\right) \text{cos}(\theta-\phi_s), 
\end{equation}
to the PIEMD potential, where $r_{s}$ is a new scale radius, $\beta$ sets the radial dependence of the skew, and $\phi_s$ is the skew angle.\footnote{The inverse tangent form of the radial dependence is not physically motivated, and other functional forms may also work. 
While it is mathematically unnecessary to have two degenerate parameters $r_s$ and $\beta$, to avoid computational divisions by zero, the distributed {\sc Lenstool} implementation uses hardcoded $r_s=0.1\arcsec$ and allows $\beta$ to vary.}
This is now available as potential 812 in {\sc Lenstool}.

The resulting surface mass densities are shown in Figure~\ref{fig:weight_skew}.
The qualitative shape of the isodensity contours changes inside or outside the scale radius (owing to the sign change in second derivative of the inverse tangent).
This feature could be used to isolate a behaviour that best matches numerical simulations, by fixing very large or very small $r_s$, or to capture more complex behaviour.

The total mass of a wPISP is identical to that of of a PIEMD.
Since the weighting function is normalised by construction, the integrated mass density of a PIEMD and wPISP over a circular region are the same:
\begin{equation}
\frac{c^2}{8 \pi G} \frac{D_s}{D_l D_{ls}}\int_{\mid \bold{r} \mid < R} \nabla ^2 \psi \text{ d}A 
=\frac{c^2}{8 \pi G} \frac{D_s}{D_l D_{ls}}\int_{\mid \bold{r} \mid < R} \nabla ^2 \left( w \psi \right) \text{ d}A,
\end{equation}
where $\theta$-dependence cancels.
Taking the limit as $\mid \bold{r} \mid \rightarrow \infty$, the left hand side will converge to the total mass of a PIEMD with ellipticity $\epsilon$, and the right to the mass of an equivalent wPISP.
However, the position of the density peak varies slightly as a function of $s$.
Care would need to be taken if using a wPISP to measure offsets of dark matter.

As with the PISP, this model breaks down for large values of $s$, since the weighting function has been applied to the potential and not the density.
We have found that the value of $s$ where this occurs is sensitive to $\beta$ and the cut and core radii.
For this reason, we recommend testing the boundaries of the parameter space for a breakdown of the desired skewed behaviour before substantial future work. 
Nonetheless, we tested the wPISP against the null hypothesis of the unskewed \emph{example\_with\_images} system distributed with {\sc Lenstool} (see \S\ref{sec:tests}). 
Fixing $\beta = 0.01$ and starting with a flat prior $s \in [0.3,0.3]$, {\sc Lenstool} successfully recovers skewness $s=0.002^{+0.002}_{-0.002}$.

\subsection{Pseudo Isothermal Varying Ellipticity Mass Distribution}

\cite{despali2016look} predict that, even with standard cold dark matter, the ellipticity of a cluster scale halo should change as a function of radius, becoming more elongated further from the centre.
This prediction can be tested by using the weighting function formalism to add an extra parameter to halo models that mimics this behaviour.
To achieve this, we combine a weighted sum of two different mass densities with different ellipticities into a Pseudo Isothermal Varying Ellipticity Mass Distribution (PIVEMD)

We suggest a mass density of the following form:
\begin{equation}
\Sigma \left( \bold{r}\right) = \Sigma_{\epsilon_1} \left( \bold{r}\right) w_1\left( \bold{r} \right) + \Sigma_{\epsilon_2} \left( \bold{r}\right) w_2\left( \bold{r} \right)
\end{equation}
where $\Sigma_{\epsilon_1} \left( \bold{r}\right)$ and $\Sigma_{\epsilon_2} \left( \bold{r}\right)$ are two elliptical profiles with ellipticity $\epsilon_1$ and $\epsilon_2$.
All the other parameters for these two densities should be shared.
In this case, we find it most effective (and possible) to apply the weighting function directly to the mass density.
To be computationally efficient within an MCMC loop, deflection angles must be computed once, using numerical integration, and stored in a look-up table.

\begin{figure} 
\centering
\includegraphics[width=0.62\linewidth]{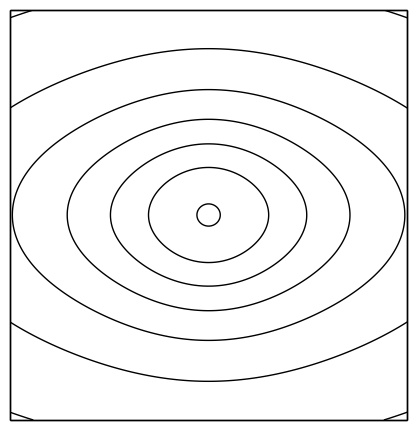}
\caption{Isodensity contours for a radially varying weighted sum of two PIEMDs with different ellipticities.}
\label{fig:vary_ellipticity}
\end{figure}

The weighting functions $w_i$ should meet the following criteria:
\begin{itemize}
\item To normalise the total mass, $w_1 \left( \bold{r}\right) + w_2 \left( \bold{r}\right) \equiv 1$, $\forall r$.
\item So that one ellipticity dominates at small $r$ and the other at large $r$, let $w_{1} \left( \bold{r} \right) \rightarrow 1 \text{ as } r \rightarrow \infty$, $w_{1} \left( \bold{r} \right) \rightarrow 0 \text{ as } r \rightarrow 0$, $w_{2} \left( \bold{r} \right) \rightarrow 0 \text{ as } r \rightarrow \infty$ and $w_{2} \left( \bold{r} \right) \rightarrow 1 \text{ as } r \rightarrow 0$. 
\end{itemize}
\par Although this is quite a general set of conditions, we can take, for example
\begin{equation}
\begin{aligned}
w_1 \left( \bold{r} \right) =& 1 - e^{- \beta r }\\
w_2 \left( \bold{r} \right) =& e^{- \beta r},
\end{aligned}
\end{equation}
where $\beta$ controls the radial dependence. 
The resulting mass distribution for this weighted sum is illustrated in Figure~\ref{fig:vary_ellipticity}.

\label{lastpage}

\end{document}